\documentstyle[aps,preprint,epsf,tighten]{revtex}
\begin{document}
\draft
\title{Phaseshift equivalent NN potentials and the deuteron}
\author{A.\ Polls}
\address{Departament d'Estructura i Costituents de la Mat\`eria,
         Universitat de Barcelona, E-08028 Barcelona, Spain}
\author{H.\ M\"{u}ther}
\address{Institut f\"ur Theoretische Physik, Universit\"at T\"ubingen,
         D-72076 T\"ubingen, Germany}
\author{R.\ Machleidt}
\address{Department of Physics, University of Idaho, Moscow,
         ID 83843, U.S.A.}
\author{M.\ Hjorth-Jensen}
\address{Nordita, Blegdamsvej 17, DK-2100 K\o benhavn \O, Denmark}
\maketitle
\begin{abstract}
Different modern phase shift equivalent NN potentials are tested by evaluating
the partial wave decomposition of the kinetic and potential energy of the
deuteron. Significant differences are found, which are traced back to the 
matrix elements of the potentials at medium and large momenta. The influence of
the localisation of the one-pion-exchange contribution to these potentials is
analyzed in detail.
\end{abstract}


During the past few years, considerable progress has been made in constructing
realistic models for the nucleon-nucleon (NN) interaction. Several of these 
models describe an
identical data base of NN scattering data with a $\chi^2$ per datum $\approx$ 1
\cite{nim,v18,cdbonn}, meaning in turn
that the on-shell matrix elements of the NN
transition matrix $T$ are essentially equal. This does, however, not imply
that the models for the NN interaction underlying these descriptions are
identical. Moreover, the off-shell properties of each potential
may be rather different. 
All models for the NN interaction $V$
include a one-pion exchange (OPE) term, using essentially 
the same $\pi NN$ coupling constant, and account for the difference between the
masses of the charged ($\pi_\pm$) and neutral ($\pi_0$) pion. 
However, even this
long range part of the NN interaction, which is believed to be well
understood, is treated quite differently in these models. As we will discuss
below, the local approximation for the OPE contribution, employed in the
non-relativistic 
Nijmegen \cite{nim} and Argonne \cite{v18} potentials, yields
significantly different contributions to the NN potential $V$  as
compared to the one derived in the relativistic framework
underlying the Bonn potential
\cite{cdbonn}. The description of the short-range part is also different in
these models. 
The NN potential
Nijm-II \cite{nim} is a  purely local potential in the sense that 
it uses the
local form of the OPE potential for the long-range part and
parametrizes the contributions of medium and short-range in terms of local
functions (depending only on the relative displacement between the 
two interacting nucleons)
multiplied by a set of spin-isospin operators. 
The same is true for the
Argonne $V_{18}$ potential \cite{v18}. The NN potential denoted by Nijm-I
\cite{nim} uses also the local form of OPE
but includes a $\bf p^2$ term in the
medium- and short-range 
central-force
(see Eq.\ (13) of Ref.\ \cite{nim})  
which may be interpreted as a non-local contribution to the central
force.
The CD-Bonn
potential is based consistently upon relativistic meson field theory
\cite{rup}. Meson-exchange Feynman diagrams are typically nonlocal expressions
that are represented in momentum-space in analytic form.
It has been shown \cite{cdbonn} that  ignoring the non-localities in the 
OPE part leads 
to a larger tensor component in the bare potential.
This is the
origin of the fact that the CD-Bonn potential yields a smaller D-state
probability for the deuteron ($P_D = 4.83 \%$) as compared to
other potentials ($P_D \approx 5.6 \%$). 

The aim of this note is to explore
characteristic features of the various phase-shift 
equivalent NN potentials depending on the model
constraints pertinent to the various potentials, such as the 
non-locality of the OPE part of the NN interaction as well as in the 
exchange terms of the heavier mesons. We have here chosen the simplest
possible system to study these differences, namely the deuteron.
Investigations along these lines, using the abovementioned
NN interactions, have been
made by studying the properties of the triton \cite{nogga}, the optical potential
for nucleon-nucleus scattering \cite{wep}, and the symmetry energy in nuclear
matter \cite{engv}. Significant differences between non-local and local NN
potentials have been observed in the study of the symmetry energy. 
As a side-remark, we ought to state that our aim, and that of the above
studies,  is different from earlier studies of phase-shift 
equivalent NN potentials, like e.g., Sauer {\it et al.} \cite{sauer,bara,haftel}. 
In those studies,
NN scattering matrices were constructed to be phase-shift equivalent
from a mathematical point of view, that is, the structure of the 
operators which determined the model for the potentials were
kept fixed, while potential parameters were allowed to vary. 

By construction,
all realistic NN potentials reproduce the experimental value for the 
energy of the deuteron of --2.224 MeV. 
However, the various contributions 
to the total deuteron energy 
originating from kinetic energy and potential energy in the
$^3S_1$ and $^3D_1$ partial waves of relative motion,
\begin{eqnarray}
   E & = & \langle \Psi_S \vert T \vert \Psi_S\rangle +  
            \langle \Psi_D \vert T \vert \Psi_D\rangle +
           \langle \Psi_S\vert V \vert \Psi_S\rangle + 
           \langle \Psi_D\vert V \vert \Psi_D\rangle + 2
           \langle \Psi_S\vert V \vert \Psi_D\rangle \nonumber \\
     & = & T_S + T_D + V_{SS} + V_{DD} + V_{SD}, \label{eq:split}
\end{eqnarray}
exhibit quite different results. This can be seen from the numbers listed in 
the upper part of 
Table \ref{tab:tab1}. In that table, we
display the various contributions to the deuteron binding 
energy employing the four
potentials introduced above.  For a comparison we also list in this table
results obtained from older NN potentials. 

\begin{figure}
\epsfysize=9.0cm
\begin{center}
\makebox[16.0cm][c]{\epsfbox{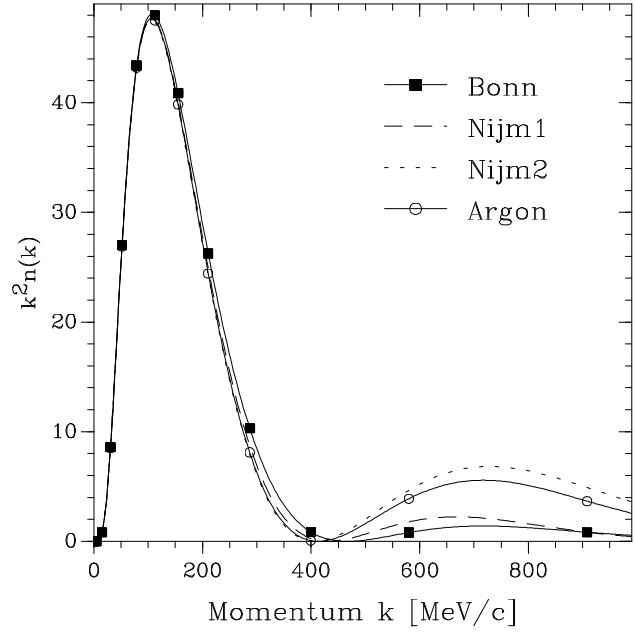}}
\end{center}
\caption{Momentum distribution in the $S$-wave of the deuteron derived from
various NN interactions. To enhance the high momentum components the
distribution has been multiplied by $k^2$ as outlined in
Eq.\ (\protect\ref{eq:mom}).}
\label{fig:wave}
\end{figure}

The kinetic energies are significantly larger for the local potentials
$V_{18}$ and Nijm II than for the two interaction models CD-Bonn and Nijm I
which contain non-local terms. 
The corresponding differences in the $S$-wave functions 
can be seen in 
Fig.\ \ref{fig:wave}, 
where 
the momentum distribution multiplied by the square of the momentum $k$ is
plotted for the $S$ wave component of the wave function
\begin{equation}
     k^2 n_S (k) = k^4 \Psi_S^2 (k).
\label{eq:mom}
\end{equation}
Note that, besides a factor $1/M$ with $M$ the nucleon mass, this function 
corresponds to the integrand for evaluating the kinetic energy $T_S$.

\begin{figure}
\epsfysize=9.0cm
\begin{center}
\makebox[16.0cm][c]{\epsfbox{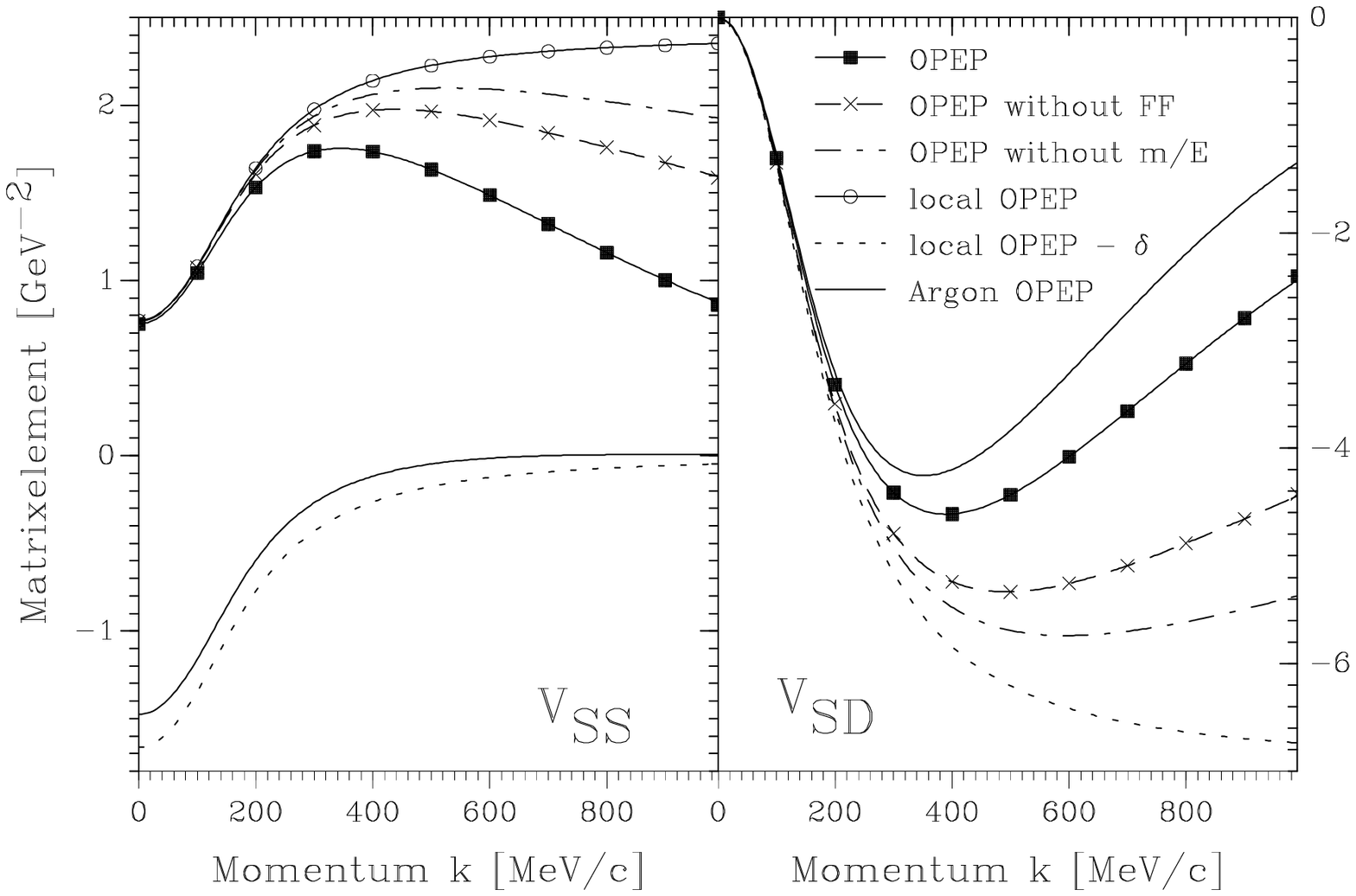}}
\end{center}
\caption{Plane-wave matrix elements of the one-pion-exchange potential 
(OPEP) using various
approximations. 
As an example, the matrix elements in momentum space 
$\langle k'\vert V\vert k\rangle$ 
are shown as functions of $k$ for a fixed value of $k'$ = 95 MeV/c.
The left part of the figure exhibits matrix elements for the partial waves
$^3S_1$--$^3S_1$, while the right part shows the tensor component in the
$^3S_1$--$^3D_1$ channel with $k$ refering 
to the momentum in the $^3D_1$ partial
wave.}
\label{fig:vpi}
\end{figure}

Comparing the contributions to the potential energy, 
one finds large differences
particularly for the tensor contribution $V_{SD}$. The dominant part of this 
tensor contribution should originate from the tensor component of the OPE
potential. This can also be seen from the last three lines of Table
\ref{tab:tab1} where the expectation value of 
the OPE part of the NN interaction
is calculated with the deuteron wave function derived from the Bonn
potential. These OPE contibutions to $V_{SS}$, $V_{DD}$ and $V_{SD}$ have been
evaluated using three different approximations to the OPE potential. 
The first one denoted by ``$\pi$
Bonn'' in Table \ref{tab:tab1} 
corresponds to the relativistic nonlocal OPE of the CD-Bonn
approach, in which the matrix elements in the
$^3S_1$--$^3S_1$ channel between plane-wave states of momenta $k'$ and $k$ 
are given by
\begin{equation}
        \langle k'\vert V_{SS}^\pi \vert k \rangle = 
         -\frac{g_\pi^2}{4\pi}\frac{1}{2\pi M^2}
         \sqrt{\frac{M^2}{E_kE_{k'}}} \int_{-1}^1 d \cos \theta 
          \left( \frac{\Lambda^2 -m_\pi^2}{\Lambda^2 + q^2}\right)^2 
         \frac{k'k\cos\theta - (E_kE_{k'}-M^2)}{q^2 + m_\pi^2},
\label{eq:vpi1}
\end{equation}
where $q^2$ denotes the momentum transfer,
\begin{eqnarray}
        q^2 & = & ({\bf k}-{\bf k'})^2 = k^2 + 
                {k'}^2 - 2 k k' \cos\theta,\nonumber \\
        E_k & = & \sqrt{k^2 + M^2},
\label{eq:momtra}
\end{eqnarray}
and 
$m_\pi$ is the mass of the pion 
(note that in order to shorten the notation we 
ignore in these equations the charge dependence of $m_\pi$ 
and the nucleon mass 
$M$); $\Lambda$ stands for the cut-off parameter, which was chosen to be
$\Lambda$ = 1.7 GeV \cite{cdbonn}. Results for such matrix 
elements as a function
of $k$, keeping $k'$ = 95 MeV/c fixed, are displayed 
in Fig.\ \ref{fig:vpi}. 
If we introduce now the nonrelativistic approximation for
\begin{equation}
     E_kE_k' - M^2 \approx \frac12 k^2 + \frac12 {k'}^2,
\label{eq:rel}
\end{equation}
we obtain
(ignoring the form factor $(\Lambda^2 -m_\pi^2)/(\Lambda^2 + q^2)$
and the relativistic square-root factors)
\begin{equation}
      \langle k'\vert V_{SS}^\pi \vert k \rangle_{\mbox{local}} = 
      - \frac{g_\pi^2}{4\pi}\frac{1}{2\pi M^2}\int_{-1}^1 d \cos \theta 
        \left(\frac{m_\pi^2}{2(q^2+m_\pi^2)} - 
        \frac{1}{2}\right). 
\label{eq:vpi2}
\end{equation}
This can be viewed as the local approximation to 
the OPE since the matrix element
depends on the momentum transfer $q^2$, only. It can easily 
be transformed into
the configuration-space representation, resulting in a 
Yukawa term plus a $\delta$
function, which originates from the Fourier transform of the constant $1/2$ in
Eq.\ (\ref{eq:vpi2}). The various steps leading to 
this result are shown in  
Fig.\ \ref{fig:vpi}. It is obvious from this figure that 
all of the steps leading
to the local expression (\ref{eq:vpi2}) are not really justified 
for momenta $k$
around and above 200 MeV, a region of relative momenta which 
is of importance in
the deuteron wave function (see Fig.\ \ref{fig:wave}).  
This is true for the matrix elements $V_{SS}$ as well as $V_{SD}$. It is
remarkable that the regularization of the local OPE by the local Gaussian
form factor introduced in the Argonne $V_{18}$ 
potential leads to matrix elements in
the $SD$-channel which are close to those derived from the relativistic
expression of the Bonn potential. This is not the case in the $SS$ channel, 
where the removal of the $\delta$ function term is very significant. This
comparison of the various approximations to the OPE part of the NN interaction
demonstrates that even this long range part of the NN 
interaction is by no means settled.
The local approximation and the 
regularization by form factors
have a significant effect.

\begin{figure}
\epsfysize=9.0cm
\begin{center}
\makebox[16.0cm][c]{\epsfbox{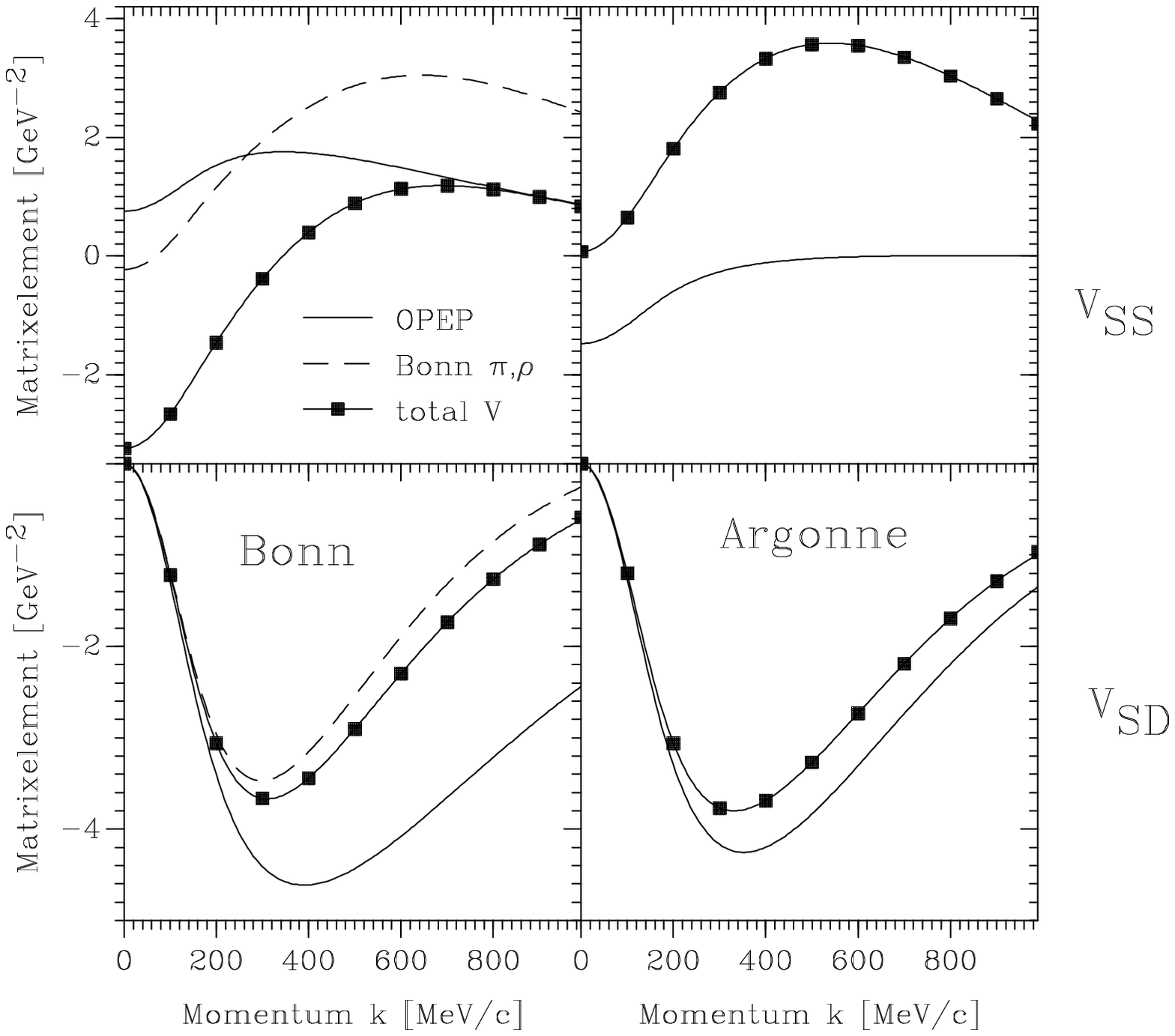}}
\end{center}
\caption{Plane-wave matrix elements of the one-pion-exchange potential 
(OPEP, solid lines)
are compared to those for the total interaction (solid lines with squares) for
the CD-Bonn potential (left part) and the Argonne $V_{18}$ potential 
(right part)
in the $^3S_1$--$^3S_1$ (upper part) and $^3S_1$--$^3D_1$ 
partial waves (lower part). For the
Bonn potential also the result of $\pi$ plus $\rho$ exchange (dashed line) is
shown. For the definition of the matrix elements 
see Fig.\ \protect\ref{fig:vpi}.
}
\label{fig:pitov}
\end{figure}

As a next step we would like to discuss the importance of the OPE contribution
as compared to the total NN interaction in the partial waves relevant for the
deuteron. As an example we consider the meson-exchange model of the Bonn
potential (figures in the left column of Fig.\ \ref{fig:pitov}) and the local
Argonne $V_{18}$ potential considering matrix 
elements in momentum space between
$^3S_1$--$^3S_1$ (upper part) and $^3S_1$--$^3D_1$ partial waves (lower part). 
The predominant
modification of the $\pi$-exchange in the OBE model is caused by 
the exchange of
the $\rho$-meson. This leads to a considerable reduction of the 
tensor component
(see $V_{SD}$ in the lower part of Fig.\ \ref{fig:pitov}) 
and enhancement of the
central contribution (see $V_{SS}$). 
The effect of other mesons is very weak in
the tensor channel and leads to a global attraction for the central part of the
NN interaction. The modification of the OPE part in the Argonne potential is
quite different. The tensor part of the NN interaction 
is reduced much less than
in the OBE model. In the central $SS$ channel the short range components of the
$V_{18}$ potential introduce considerable repulsion 
but lead to a momentum dependence which, apart
from a repulsive constant, is similar to that of the Bonn potential.

\begin{figure}
\epsfysize=9.0cm
\begin{center}
\makebox[16.0cm][c]{\epsfbox{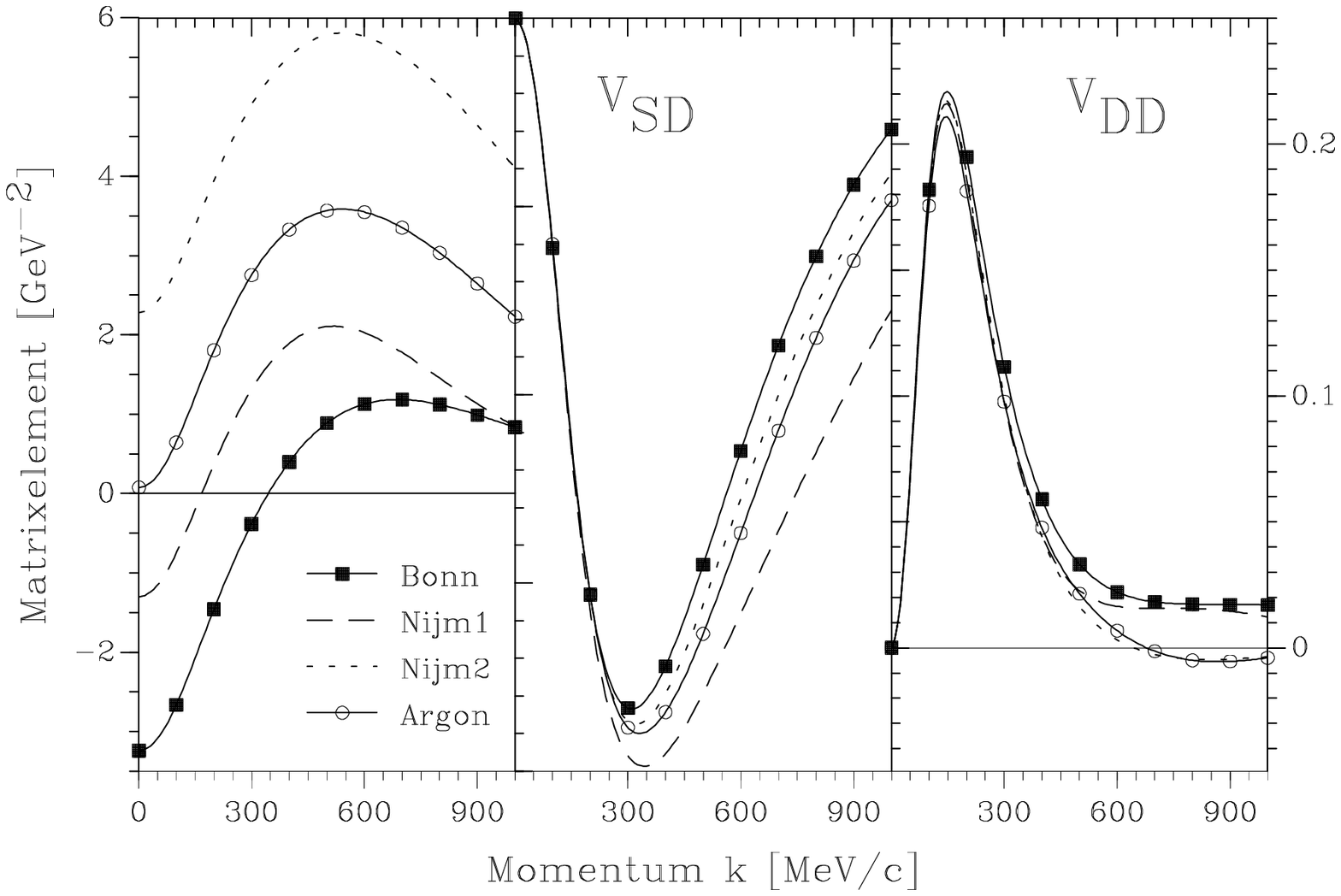}}
\end{center}
\caption{Plane-wave matrix elements of various realistic NN 
interactions in momentum space
for the partial wave channels $^3S_1$--$^3S_1$, $^3S_1$--$^3D_1$, and 
$^3D_1$--$^3D_1$. For the definition of the matrix elements see Fig.\ 
\protect\ref{fig:vpi}.}
\label{fig:vcomp}
\end{figure}

Finally, we compare in Fig.\ \ref{fig:vcomp} 
the momentum space matrix elements of
all four phase-shift equivalent charge-dependent NN potentials. Substantial
differences are observed in the central $SS$ channel as well as in the tensor
channel $SD$. The local approximation of the OPE contribution to the tensor
term is highly suppressed at large momenta (compare Fig.\ \ref{fig:vpi}). The
differences between the local approximation of the OPE tensor term and the
nonlocal treatment shown in Fig.\ \ref{fig:vpi} is much larger than the
final differences between the various potentials shown 
in Fig.\ \ref{fig:vcomp}.
In the OBE model the final shape of the tensor potential  is achieved 
by means of relativistic corrections, form factors, and $\rho$ exchange. The
local potentials describe the reduction of the local OPE mainly in terms 
of local form factors. From the remaining differences we see that the 
requirement of identical NN  scattering phase shifts
provides only a moderate constraint on this tensor component. As a consequence 
one obtains different D-state probabilities in the deuteron 
wave function (see
Table \ref{tab:tab1}). It seems that the range of 
possible D-state probabilities
has been reduced to some extent by the improvements in NN scattering data
analysis (as compared to the range produced by the older potentials). 

The momentum dependence of the various NN potentials in the 
central $SS$ channel
is rather similar. Here the main differences can be described in terms of a
constant in momentum space or a corresponding 
$\delta$ function in configuration
space. It is worth mentioning that the two non-local potentials Nijm I and 
CD-Bonn 
yield matrix elements, which are closer to each other as compared to
the other two potentials.
The same is true for the $V_{DD}$ matrix elements. This may also be the origin 
for the differences in the high momentum components of the wave functions
(cf.\ Fig.\ \ref{fig:wave}).

In summary, although the modern NN potentials yield the same
binding energy for the deuteron, there are significant differences 
in the contributions to both the kinetic and potential energy
in the various partial waves. Speaking in general terms, these    
differences can be traced back to 
off-shell differences between the potentials.
It is well known that the off-shell behavior of NN potentials
cannot be pinned down by on-shell data.
On the other hand, within realistic physical models, not everything
is possible that mathematics may allow.
Moreover, within a given model, pinning down the on-shell point
limits the range of variation for the off-shell behavior.
Therefore, since the four models considered in this investigation
reproduce the NN scattering data with the perfect $\chi^2$/datum
$\approx 1$, and assuming that the differences between the models
cover about the range of diversity that there is to realistic
physicial models for the NN interaction, one may conclude
that the off-shell uncertainties revealed in this study cannot be
further reduced by objectively verifiable aspects.
Further constraints for the off-shell behavior of NN potentials,
if any, may come from our preference for a particular theory
or from the many-body problem.

This project has been  supported from the EC-program `Human
Capital and Mobility' under Contract N. CHRX-CT 93-0323 and by the DGICYT
(Spain) Grant PB95-1249.
One of the authors (R.M.) acknowledges partial support
by the US National Science Foundation
under Grant No.\ PHY-9603097.

\begin{table}
\begin{tabular}{| c | r r r r r r |}
Pot. & $T_S$ & $T_D$ & $V_{SS}$ &  $V_{DD}$ & $V_{SD}$ & $P_D$ \\
& [MeV] & [MeV] &[MeV] &[MeV] &[MeV] & [ \% ] \\
\hline
&&&&&&\\
CD-Bonn & 9.79 & 5.69 & -4.77 & 1.34 & -14.27 & 4.83 \\
Argon $V_{18}$ & 11.31 & 8.57 & -3.96 & 0.77 & -18.94 & 5.78 \\
Nijm I & 9.66 & 7.91 & -1.35 & 2.37 & -20.82 & 5.66 \\
Nijm II & 12.11 & 8.10 & -5.40 & 0.59 & -17.63 & 5.64 \\
&&&&&&\\
\hline
&&&&&&\\
Bonn A & 10.05 & 4.41 & -7.29 & 0.39 & -9.78 & 4.38 \\
Bonn B & 10.02 & 5.62 & -5.39 & 1.01 & -13.49 & 4.99 \\
Bonn C & 10.13 & 7.44 & -1.20 & 3.46 & -22.05 & 5.62 \\
Argon $V_{14}$ & 10.53 & 8.68 & -1.83 & 1.99 & -21.59 & 6.08 \\
Reid & 12.61 & 9.53 & -0.47 & 4.55 & -28.45 & 6.47 \\
Urbana & 11.15 & 6.44 & -6.91 & -0.19 & -12.73 & 5.20 \\
&&&&&&\\
\hline
&&&&&&\\
$\pi$ Bonn & & & 2.32 & 1.86 & -21.48 & \\
$\pi$ local & & & -1.99 & 2.18 & -26.76 & \\
$\pi$ Argon & & & -1.65 & 1.60 & -18.28 & \\
&&&&&&\\
\end{tabular}
\caption{Contributions to the kinetic and potential energy 
of the deuteron originating from the 
$^3S_1$
and $^3D_1$ parts of the wave function as defined 
in Eq.\ \protect\ref{eq:split}. 
Results are
listed for the charge-dependent Bonn potential (CD-Bonn \protect\cite{cdbonn}),
the Argonne $V_{18}$ \protect\cite{v18}, the 
Nijmegen potentials Nijm I and Nijm
II \protect\cite{nim}, the Bonn potentials A,B,C \protect\cite{rup}, 
the Argonne
$V_{14}$ \protect\cite{v14}, the Reid soft-core potential \protect\cite{reid}
and the Urbana potential \protect\cite{urban}. Also listed are the expectation
values of the charge-dependent OPE part, using the non-local 
approach from the CD-Bonn potential (`$\pi$ Bonn'),
the local approximation (`$\pi$ local'), and the local approximation 
with inclusion of the
Argonne form factor (`$\pi$ Argon'). 
In the latter three cases, these expectation values are
calculated by employing 
the CD-Bonn deuteron
wave function. The last column of this Table shows the calculated $D$-state
probabilities of the deuteron.} 
\label{tab:tab1}
\end{table}

\vfil\eject
\vfil\eject

\vfil\eject

\end{document}